\documentclass[runningheads]{llncs}
\usepackage{graphicx}
\usepackage{amsmath,graphicx}
\usepackage{url}
\usepackage{booktabs}
\usepackage{cite}

\newcommand{\etal}{\textit{et al.\,}}
\newcommand{\eg}{\textit{e.g.}}
\newcommand{\ie}{\textit{i.e.}}

\usepackage{xcolor}
\usepackage{bbding}
\usepackage{pifont}

\usepackage{cleveref}
\crefname{appendix}{App.\negthinspace\,}{App.\negthinspace\,}
\crefname{chapter}{Chap.\negthinspace\,}{Chap.\negthinspace\,}
\crefname{equation}{Eq.\negthinspace\,}{Eq.\negthinspace\,}
\crefname{algorithm}{Alg.\negthinspace\,}{Alg.\negthinspace\,}
\crefname{section}{Sec.\negthinspace\,}{Sec.\negthinspace\,}
\crefname{subsection}{Sec.\negthinspace\,}{Sec.\negthinspace\,}
\crefname{subsubsection}{Sec.\negthinspace\,}{Sec.\negthinspace\,}
\crefname{figure}{Fig.\negthinspace\,}{Fig.\negthinspace\,}
\crefname{table}{Tab.\negthinspace\,}{Tab.\negthinspace\,}
\crefname{subfigure}{Fig.\negthinspace\,}{Fig.\negthinspace\,}
\crefname{subsubfigure}{Fig.\negthinspace\,}{Fig.\negthinspace\,}
\crefname{lstlisting}{Lst.\negthinspace\,}{Lst.\negthinspace\,}

\begin{document}
\title{Towards Automatic Embryo Staging in 3D+t Microscopy Images using Convolutional Neural Networks and PointNets\thanks{We thank the Helmholtz Association in the program BIFTM (MT), the German Research Foundation DFG (JS, Grant No STE2802/1-1) and the colleagues M.~Takamiya, A.~Kobitski, G.~U.~Nienhaus, U.~Str\"ahle and R.~Mikut at the Karlsruhe Institute of Technology for providing the microscopy data and for the collaboration on previous analyses that form the basis of the data analyzed in this work.}}
\titlerunning{Towards Automatic Embryo Staging in 3D+T Microscopy Images}
\author{Manuel Traub\inst{1,2}\orcidID{0000-0003-0897-1701} \and Johannes Stegmaier\inst{1}\orcidID{0000-0003-4072-3759}}

\authorrunning{M. Traub et al.}
%\authorrunning{X. XXXXX et al.}
\institute{Institute of Imaging and Computer Vision, RWTH Aachen University, Germany \and Institute for Automation and Applied Informatics, Karlsruhe Institute of Technology, Germany\\
%\email{johannes.stegmaier@lfb.rwth-aachen.de}
%\institute{XXX \and XXX}\\
\email{johannes.stegmaier@lfb.rwth-aachen.de}}

%\email{xxxxxxxx.xxxxxxxxx@xxx.xxx}}
%
\maketitle              % typeset the header of the contribution
\begin{abstract}
Automatic analyses and comparisons of different stages of embryonic development largely depend on a highly accurate spatiotemporal alignment of the investigated data sets. In this contribution, we assess multiple approaches for automatic staging of developing embryos that were imaged with time-resolved 3D light-sheet microscopy. The methods comprise image-based convolutional neural networks as well as an approach based on the PointNet architecture that directly operates on 3D point clouds of detected cell nuclei centroids. The experiments with four wild-type zebrafish embryos render both approaches suitable for automatic staging with average deviations of $21 - 34$ minutes. Moreover, a proof-of-concept evaluation based on simulated 3D+t point cloud data sets shows that average deviations of less than $7$ minutes are possible.

\end{abstract}
\section{Introduction}
\label{sec:intro}
\begin{figure*}[h]
  \includegraphics[width=\textwidth]{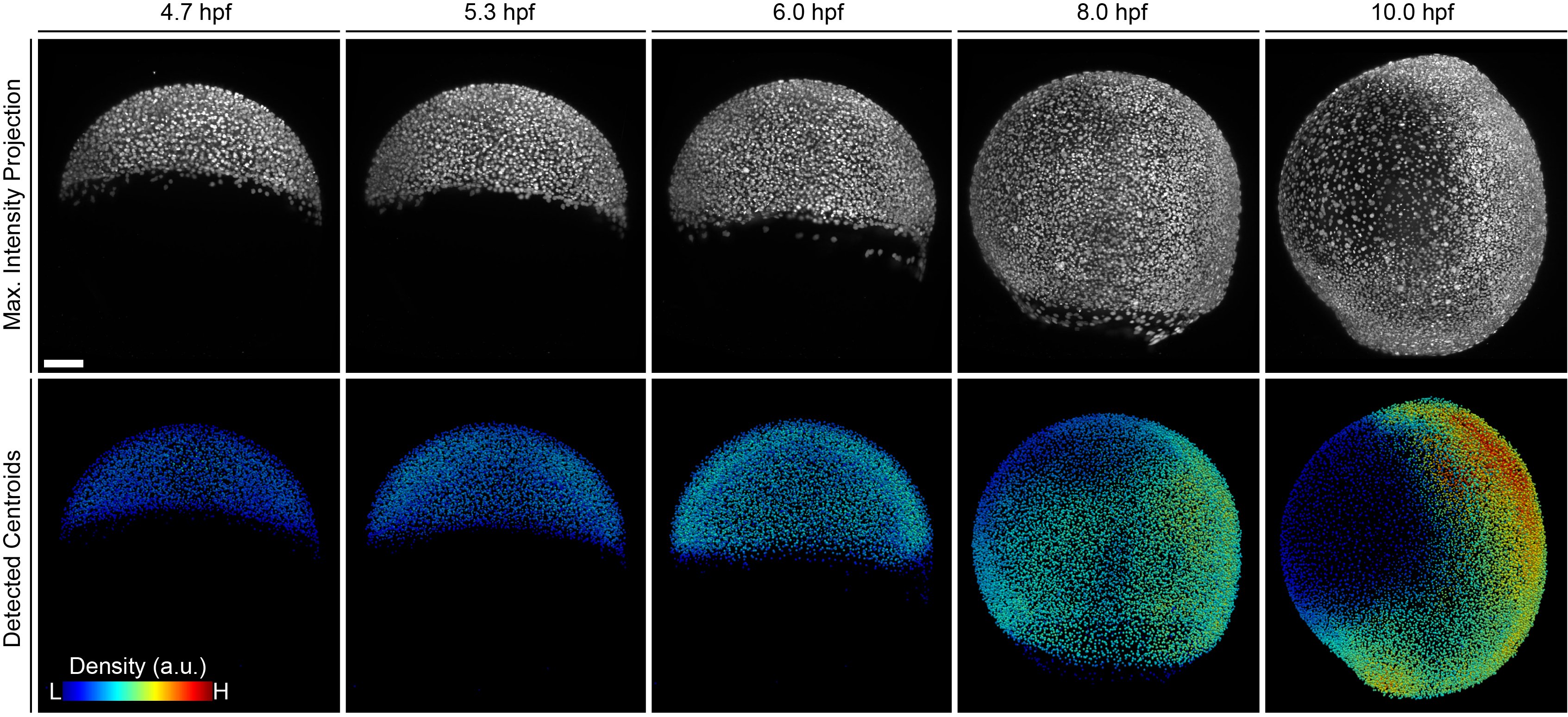}
  %\vspace{-0.7cm}
  \caption{Maximum intensity projections of 3D light-sheet microscopy images of a zebrafish embryo imaged from 4.7 - 10 hpf that were used as the input for the CNN-based stage predictions (top). Centroids of the fluorescently labeled cell nuclei were automatically detected \cite{Stegmaier14b} and are visualized with a density-based color-code that measures the relative number of neighbors in a fixed sphere neighborhood (bottom). A fixed number of points sampled from the centroids of different time points directly serves as input to the PointNet-based stage prediction. Scale bar: $100\mu m$.}
  \vspace{-0.3cm}
  \label{fig:Figure1}
  \end{figure*}
Embryonic development is characterized by a multitude of synchronized events, cell shape changes and large-scale tissue rearrangements that are crucial steps in the successful formation of a new organism \cite{Lecuit07}. To be able to compare these developmental events among different wild-type individuals, different mutants or upon exposure to certain chemicals, it is highly important to temporally synchronize acquired data sets, such that corresponding developmental stages are compared to one another \cite{Castro-Gonzalez14,Guignard14}. 
While the temporal synchronization is typically easy in small 2D screens, it becomes already a tedious undertaking when analyzing high-throughput screens consisting of thousands of repeats. Shifting dimensions to 3D the challenge of reproducible temporal synchronization becomes even more difficult and finally almost impossible for a human observer if the time domain is additionally considered. 
Current approaches to embryo staging largely rely on human intervention with the risk of subjective bias and might require specific labeling strategies or sophisticated visualization tools to cope with large-scale time-resolved 3D data \cite{Pietzsch15, Schott18}.

Under the assumption that development progresses equally in all embryos, a rough temporal synchronization of the data sets can be obtained by measuring the time between fertilization and image acquisition \cite{Shahid16}. Moreover, a visual staging can be performed after image acquisition by identifying certain developmental characteristics of a single snapshot of a time series or after chemical fixation of the embryo and by comparing it with a series of standardized views that show the characteristic development of a wild-type specimen at a standardized temperature \cite{Kimmel95}. In early phases of development, synchronized cell divisions regularly double the cell count and can be used to reliably align early time points \cite{Faure16,Kobitski15}. Villoutreix \etal use measured cell counts over time to specify an affine transformation on the temporal domain, \ie, the time axis is scaled such that the cell count curves of multiple embryos best overlap \cite{Villoutreix16}. While cell counts might be reliable in specimens like the nemathode \textit{C. elegans}, where adult individuals exhibit a largely identical number of cells, using the cell counts for synchronization in more complex animal models becomes more and more ambiguous. As development progresses, strong variation of total cell counts, cell sizes, cell shape and tissue formation are observed even among wild-type embryos \cite{Fowlkes08, McDole18}. If the number of individuals is limited and if later time points are considered, a manual identification of characteristic anatomical landmarks can be used to assign a specific developmental stage to selected frames \cite{Muenzing18, McDole18, Winkley18}. With steadily increasing degrees of automation of experimental setups, it will be impractical to perform the staging with the human in the loop and automated approaches could thus help to further automate experimental protocols. 

In this contribution, we analyzed two learning-based approaches for their suitability to automate embryo staging in large-scale 3D+t microscopy experiments. The methods comprise image-based convolutional neural networks (CNNs) as well as a point cloud-based approach using the PointNet architecture \cite{Qi17a}. Both approaches were adapted for regression tasks and assessed under different hyperparameter and training conditions. We applied the methods to 2D maximum intensity projections as well as to real and simulated 3D point clouds of cell nuclei centroids (\cref{fig:Figure1}). Training data was extracted from terabyte-scale 3D+t light-sheet microscopy images of four wild-type zebrafish embryos (\emph{D. rerio}) that ubiquitously expressed a green fluorescent protein (GFP) in their cell nuclei \cite{Kobitski15}.

\section{Automatic Embryo Staging as a Regression Problem}
We analyzed two conceptually different approaches to automatically predict the hours post fertilization (hpf), a common measure for staging zebrafish embryos, either from 2D maximum intensity projection images or from 3D point clouds of centroids of fluorescently labeled cell nuclei. To estimate the achievable staging accuracy to be expected for training with a perfect ground truth, we created a synthetic 3D+t point cloud data set by adapting the method described in \cite{Stegmaier16d}.  We intentionally did not use cell counts for staging due to large inter-specimen variations. Instead, the networks should learn to derive the stage solely from the appearance in the projection (CNN-based pipelines) or the arrangement of nuclei in space (PointNet-based pipeline).

\subsection{CNN-based Embryo Staging on Downsampled Maximum Intensity Projections}
\begin{figure}[h]
\begin{center}
    \includegraphics[width=0.7\columnwidth]{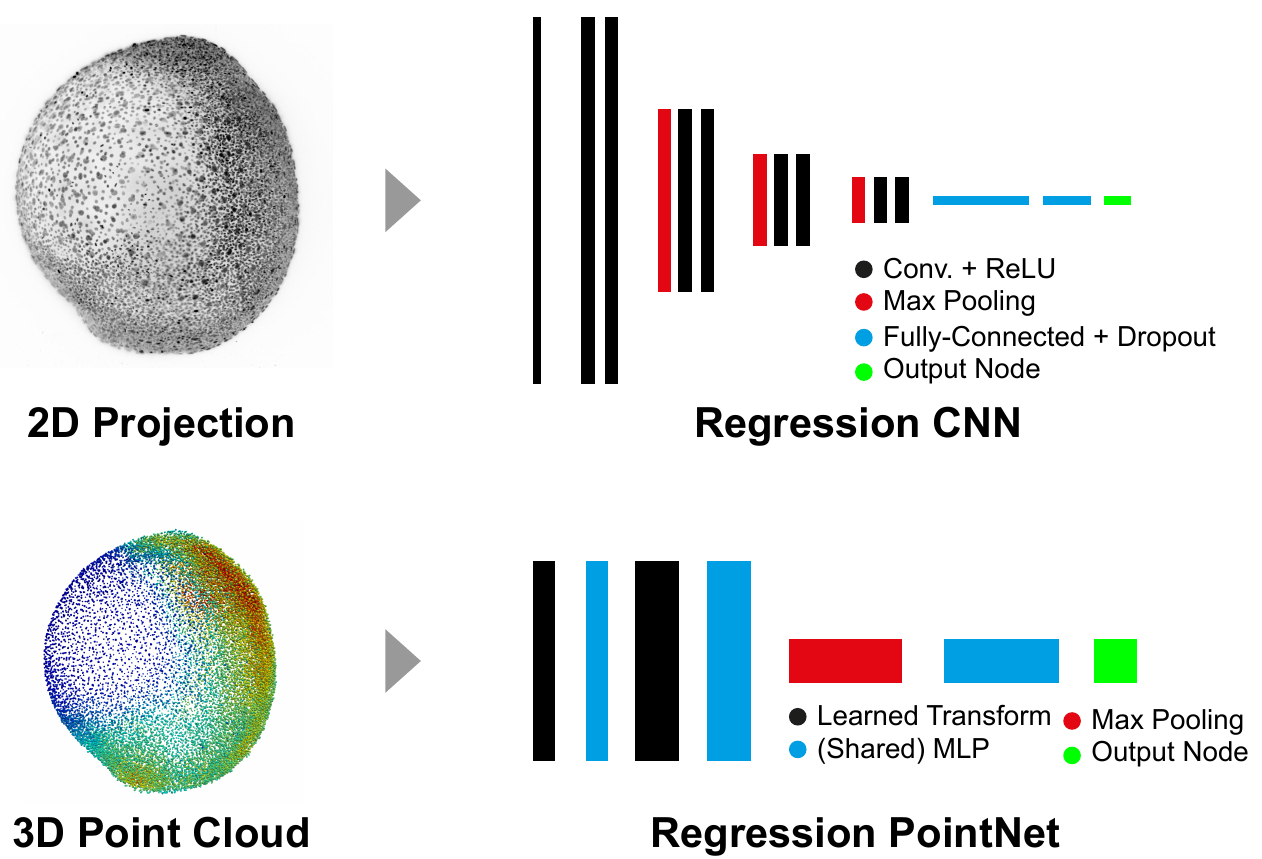}
    %\vspace{-0.7cm}
    \caption{Regression approaches used for automatic embryo staging. Different CNNs are applied to 2D maximum intensity projection images (top) and a PointNet \cite{Qi17a} is applied to 3D point clouds of detected cell nucleus centroids (bottom). The depicted regression CNN architecture (top) is referred to as VGG-Simple in the remainder of the paper.}
    \vspace{-0.3cm}
    \label{fig:Figure2}
    \end{center}
\end{figure}

For the image-based approach we selected three established CNN architectures, namely VGG-16, ResNet-18 and GoogLeNet \cite{Simonyan14,He16,Szegedy15}. The classification output layers were replaced with a single regression node with a linear activation to predict the stage of the current input in hpf. In addition to the relatively large pretrained networks, we added a more shallow VGG-like network (VGG-Simple) consisting of four blocks of convolutions followed by ReLU activation ($3 \times 3$ kernels, stride 1, 32 layers in the first convolutional layer and doubling the depth after each pooling operation), three max-pooling layers ($2 \times 2$ kernels, stride 2) and two fully-connected layers (128 and 64 nodes, each followed by a dropout layer with probability $p=0.2$) that map to the final regression node (\cref{fig:Figure2}).

%Similar behavior could be observed for batch normalization, which led to oscillating training and test results. We hypothesize that this was due to the homogeneous black background that is probably affected by batch normalization and thus possibly misleads the training process to search for distinctive information in background regions.

\subsection{Regression PointNet for Automatic Embryo Staging}

To perform the automatic staging on 3D point clouds, we equipped the original PointNet architecture by Qi \etal \cite{Qi17a} with a regression layer as the target instead of classification scores (\cref{fig:Figure2}, bottom). In brief, the PointNet architecture processes input points separately by first putting them through learnable input and feature transformation modules ($3\times3$ and $64\times64$) that are connected via a point-wise multilayer perceptrons with shared weights. This step is intended to orient the point cloud consistently and thus to obtain invariance with respect to input transformations. The transformed points are then fed through another set of multilayer perceptrons with shared weights (64, 128 and 1024 nodes in the hidden layers), a max-pooling operation that serves as a symmetric function to compensate for permutations of the input points and a final fully-connected multilayer perceptron with 512 and 256 hidden layers that map to \emph{k} output scores in the original classification PointNet. To use PointNet for regression, we change the final fully-connected layer to map to a single output neuron with linear activation and use a mean squared error loss to regress to the target measure in hpf. As the PointNet operates only on a subset of the input points during each forward pass, we use an ensemble average of the predictions over $25$ runs with randomly selected input points for the final output.

\subsection{Synthetic Embryo Data Set to Identify the Possible Accuracy}
\label{sec:Simulation}
As the temporal comparability of real embryos may be biased by the manual synchronization or due to inter-experiment variability caused by deviations from the assumption of the linear development, we decided to add an additional validation experiment using synthetic 3D+t point clouds that mimic embryonic development. As described in \cite{Stegmaier16d}, virtual agents can be placed on tracked cell centroids of a real embryo, even in the presence incomplete or erroneous tracks. However, we simplified the approach described in \cite{Stegmaier16d}, while still visually improving the realism of the simulations. The major difference is that the simulation is performed backwards in time, \ie, we start at the last time point and initialize the simulation with a fraction of $p\in\left[0, 1\right]$ randomly selected real positions at this time point. Each sampled position $\mathbf{x}_i$ was additionally modified by adding a small displacement in a random direction at most half the distance to its spatially nearest neighbor. So even if the same real cells were sampled, the additional randomization of each location ensures that no identical positions are present when creating multiple simulations of a single embryo. We skip the repulsive and nearest neighbor attraction forces and only rely on the movement directions of the closest neighbors in the real data set to update the positions of the simulated objects. The total displacement vector $\Delta \mathbf{x}^{\text{tot}}_i$ for each simulated object $i$ at location $\mathbf{x}_i$ thus simplifies to the average of the backward directions $-\mathbf{d}_j$ of its $K$ spatially nearest neighbors $j\in \mathcal{N}
^K_{\text{knn}}(\mathbf{x}_i)$ within the real embryo:

\begin{equation}
    \Delta \mathbf{x}^{\text{tot}}_i = \Delta \mathbf{x}^{\text{dir}}_i = \frac{1}{K} \sum_{j \in \mathcal{N}^K_{\text{knn}}(\mathbf{x}_i)} -\mathbf{d}_j.
\end{equation}
As we start with the maximum number of objects and perform a time-reversed simulation, we need to incorporate merge events rather than cell division events. The number of merges $N^{\text{merge}}_k$ required in time point $k$ is defined as the difference of the objects currently present in the simulation $N^{\text{sim}}_k$ and the target number of objects determined by the desired fraction of the real objects $p \cdot N^{\text{embryo}}_k$: 
\begin{equation}
    N^{\text{merge}}_k = \max \left( 0, N^{\text{sim}}_k - p \cdot N^{\text{embryo}}_k  \right).
\end{equation}
To determine which of the current objects should be merged to preserve the density distribution of the real embryo even in the case of fewer simulated objects, we adapted the relative density difference measure from \cite{Stegmaier16d} and compute it for each object $i$ at time point $k$:

\begin{equation}
    \rho^{\text{diff}}_{ik} = \frac{\rho^{\text{sim}}_{ik}}{N^{\text{sim}}_k} - \frac{\rho^{\text{embryo}}_{ik}}{N^{\text{embryo}}_k}.
\end{equation}
The density $\rho_{ik}$ was defined as the number of neighbors present in a sphere with a fixed radius of $50~\mu m$ centered at each object in the real or the simulated point cloud, respectively. Although the simulation is not sensitive to the exact value selected for the neighborhood radius, it should be large enough that even in sparse regions of the simulation at least a few neighbors are present. Moreover, it is important to use the same radius for both the real and the simulated embryo, to ensure comparable density differences after normalizing the number of neighbors with the number of total objects at each time point. We select the $N^{\text{merge}}_k$ simulated cells at time point $k$ with the maximum relative density difference $\rho^{\text{diff}}_{ik}$ and combine each of these objects with its spatially closest neighbor at time point $k$ to a single fused object at time point $k-1$. As the PointNet-based stage identification is most relevant for practical applications, we did not simulate artificial image data and only focused on a 3D+t point cloud-based scenario for the identification of the achievable accuracy for a perfect ground truth.

\section{Experiments}

\subsection{Data Acquisition and Training Data Generation}
Raw image data was acquired using a custom-made light-sheet microscope as described in \cite{Kobitski15}. Experiments considered in this contribution comprise four wild-type zebrafish embryos expressing a transgenic H2A-GFP fusion protein in their cell nuclei that were imaged from about $2.5$ to $16$ hpf and were performed in accordance with the German animal protection regulations, approved by the Regierungspr\"asidium Karlsruhe, Germany (Az. 35-9185.64) \cite{Kobitski15}. A single experiment acquired with this technique accumulates about $10$ terabytes of raw image data, \ie, the total amount of data analyzed was on the order of 40~TB. Cell nuclei of all four embryos were automatically segmented using the TWANG segmentation algorithm \cite{Bartschat16a, Stegmaier14b} and we aligned the 3D+t point clouds in a consistent orientation with the prospective dorsal side pointing to the positive x-axis, the anteroposterior axis oriented along the y-axis \cite{Kobitski15,Schott18}. A temporal window of $4.7 - 10$ hpf was selected by manually identifying the 10 hpf stages as a synchronization time point and assuming constant development prior to that stage. The time interval was selected, such that the segmentation worked reliably with sufficiently strong fluorescent signal in the early time points and the ability to still resolve single cell nuclei in high density areas in later frames \cite{Schott18}. Stage identification was performed on 3D+t point clouds visualized in KitWare's ParaView \cite{Ayachit15}. The manual stage identification required several hours of interactive data visualization and analysis, which could largely benefit from automation.

For the image-based CNNs, we computed 2D maximum intensity projections for all data sets and all time points along the axial direction. Intensity values of all frames were scaled to the 8 bit range and contrast adjusted such that $0.3$\% of the pixels at the lower and the upper end of the intensity range were saturated. For networks pretrained on the ImageNet database, we scaled the single channel 8 bit images to $224 \times 224$ pixels with three redundant channels. We ended up with $370$ frames for each embryo spanning a duration of $4.7 - 10$ hpf, \ie, $1480$ images for all embryos in total. The PointNet was trained on 3D centroids of detected nuclei, which were extracted from the same four embryos as for the image-based experiments (between $4160$ to $19794$ per data set) \cite{Schott18}.
In addition to the real data sets, we created four simulations using $p=0.75$, \ie, $75\%$ of the real detections and movements of a single embryo but with different random initializations, such that we obtained the same overall shape and the same density distribution over time. The target hpf values were identical to the ones we used for the real embryo spanning from $4.7 - 10$ hpf over $370$ frames. The achievable performance is only assessed using the regression PointNet, as this is the practically most relevant approach and also allows to assess the 3D shape of the embryo irrespective of its orientation in the sample chamber rather than a 2D projection with potential occlusions.

\subsection{Implementation and Training Details}
The CNN-based approaches and pretrained networks were implemented in MATLAB using the Deep Learning Toolbox. We used the ADAM optimizer, tried different mini-batch sizes of $8, 16, 32$ and $64$ and trained for 100 epochs (sufficient for all models to converge). Optionally, we used data augmentation including reflection, scaling ($0.9-1.0$), random rotations ($0-180^\circ$) and random translations ($\pm 5$ px). In addition to training the networks from scratch, we investigated the performance of fixing the pretrained weights of the networks at different layers and only fine-tuned the deeper layers to the new task. For VGG-16, we tested all positions prior to the max-pooling layers (5 possibilities), for ResNet-18, all positions before and after pairs of ResNet modules (5 possibilities) and for GoogLeNet, all locations between Inception modules (9 possibilities).
The PointNet was implemented using TensorFlow by modifying the original repository \cite{Qi17a}. Training was performed using the ADAM optimizer and we randomly sampled $4096$ centroids of each time point at each training iteration as the fixed-size input to the PointNet. We consistently obtained better results when disabling dropout and batch normalization during training and thus only report these results. For augmentation, random rotations around the origin were applied to all points and point jittering was used to apply small random displacements to each of the 3D input points. Training was performed with 4-fold cross-validation using three embryos for training and one embryo for testing in each fold, respectively. Simulated embryos were created using MATLAB and processed like the real data. Reported average values and standard deviations were computed on all folds.

\section{Results and Discussion}
\begin{figure*}[t]
    \includegraphics[width=\textwidth]{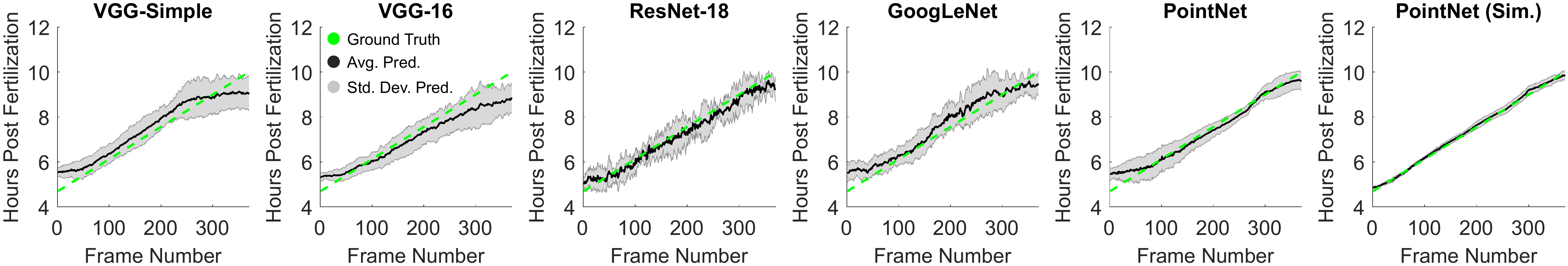}
    %\vspace{-0.7cm}
    \caption{Four-fold cross-validation results for the CNN-based predictions on maximum intensity projections (VGG-Simple, VGG-16, ResNet-18 and GoogLeNet, respectively) and the results of the regression PointNets (right). Ground truth is depicted in green and the mean $\pm$ std.\,dev.\,of the four cross-validation runs are depicted with black lines and gray shaded areas. Note that the PointNet (Sim.) approach was trained on synthetic data obtained from a single embryo to approximate the achievable accuracy.}
    \vspace{-0.3cm}
    \label{fig:Figure3}
    \end{figure*}

The results of the best combination of architecture, augmentation and hyperparameters are summarized in \cref{tab:Performance} and \cref{fig:Figure3}. The smallest mean deviation from the ground truth was obtained with the PointNet approach with an average deviation of $0.35 \pm 0.24$ hours (mean $\pm$ std.\,dev.). ResNet-18 was the best scoring CNN-based approach with an average deviation of $0.45 \pm 0.30$ hours with a mini-batch size of $32$ and with freezing the pretrained weights of the first four ResNet modules. These results were closely followed by the GoogLeNet trained from scratch with enabled data augmentation and a mini-batch size of 8 ($0.50 \pm 0.37$ hours) and the VGG-Simple architecture ($0.51 \pm 0.40$ hours) with augmentation enabled and a mini-batch size of $16$. VGG-16 yielded the poorest performance in all investigated settings and the best results reported in \cref{tab:Performance} ($0.57 \pm 0.41$ hours) were obtained using a fine-tuning approach with keeping the weights up to the third convolutional block fixed, with enabled data augmentation and a mini-batch size of $8$. The 138 million parameters of VGG-16 combined with a limited amount of training images showed a tendency to overfit despite using data augmentation and dropout regularization. However, with fixed pretrained weights and enabled data augmentation the qualitative trend did not change and the network even failed to learn the training data properly, which is probably due to the differences of the ImageNet data from fluorescence microscopy images. The same ranking was obtained when considering the root mean square deviation (RMSD, \cref{tab:Performance}). To approximate the potentially possible accuracy, we assessed the performance of the PointNet on simulated data, reaching an average deviation of $0.11 \pm 0.09$ hours ($6.6 \pm 5.4$ min). Both in the early as well as in the late frames the major differences originate from cell density rather than shape, which is not yet considered by the models. Thus, the best performance was consistently achieved in intermediate frames ($\approx$ 50 -- 300) which are characterized by large-scale tissue rearrangements during the gastrulation phase.

The performance of the image-based projection approach directly depends on the orientation of the embryo during image acquisition. On the contrary, the point cloud-based method is essentially independent of the initial orientation and randomized orientation of the point clouds are explicitly used as augmentation strategy during the training to further improve the invariance of the stage prediction with respect to the spatial orientation. The staging can thus be performed even without a tedious manual alignment of different experiments or the use of registration approaches, which might be required for the image-based approach. Robustness with respect to specimen orientation could be achieved for the image-based CNN approach by using 3D CNNs that are directly applied on downsampled raw image stacks \cite{Qi16}. However, this would require extensive preprocessing of the terabyte-scale data sets like multi-view fusion and downsampling, which would thus dramatically increase the required processing time. Currently, the time required for predicting the stage of a single time point is on the order of milliseconds for all investigated approaches.
\begin{table}[htbp]
\begin{center}
\caption{Staging accuracy of the different methods with top-scoring methods highlighted in bold-face letters. Last row: results obtained on the simulated data sets.}\label{tab:Performance}
\vspace{-0.5cm}
\begin{tabular}{|l|c|c|c|}
\hline
\textbf{Method} & \textbf{Modality} & \textbf{Mean Dev. $\pm$ Std. Dev. (h)} & \textbf{RMSD (h)} \\
    \hline
    VGG-Simple & 2D Images & 0.51 $\pm$ 0.40 & 0.65 \\
    VGG-16 \cite{Simonyan14} & 2D Images & 0.57 $\pm$ 0.41 & 0.70 \\
    ResNet-18 \cite{He16} & 2D Images & 0.45 $\pm$ 0.30 & 0.54 \\
    GoogLeNet \cite{Szegedy15} & 2D Images & 0.50 $\pm$ 0.37 & 0.62 \\
    %PointNet \cite{Qi17a} & 3D Point Cloud & 0.47 $\pm$ 0.29 & 0.55 \\
    PointNet \cite{Qi17a} & 3D Point Clouds & \textbf{0.35 $\pm$ 0.24} & \textbf{0.42} \\
\hline
    PointNet \cite{Qi17a} & 3D Point Clouds (Sim.) &  0.11 $\pm$ 0.09 & 0.14 \\
\hline
\end{tabular}
\vspace{-0.5cm}
\end{center}
\end{table}

The general average trends obtained by PointNet, ResNet-18 and GoogLeNet nicely resemble the ground truth and are arguably the most promising candidates among the presented methods. While an average deviation of about $21-30$ minutes from the real developmental time can be readily used for a global staging and temporal alignment of long-term experiments, applications involving rapid tissue changes that potentially can happen in a matter of a few minutes might demand for a higher accuracy. However, as the experiments with simulated data shows, given sufficiently accurate ground truth data an accuracy of less than $7$ minutes becomes possible. Moreover, we also expect a higher variability of the images in phases of rapid tissue changes that could potentially improve the separability. We consider the presented methods as a first proof-of-concept for automatic embryo staging in 3D+t experiments showcasing two conceptually different approaches. To compensate the lack of training data, we used several data augmentation strategies including various image transformations, point jittering and random rotations of the entire data set. However, data augmentation in the image-space did not consistently improve the results. As a straightforward extension, we could simply enlarge the training set by additional experiments or pre-train the real networks on realistic simulations as the ones presented in \cref{sec:Simulation}. Moreover, the data sets we used for demonstration were manually synchronized and assumed constant development of the embryo prior to the synchronization time point, with a potential subjective bias. To obtain a more objective ground truth calibration experiments would be required, \eg, by imaging multiple channels of reporter genes that are known to be expressed within a distinct time window. Finally, an automatic spatial registration of the temporally aligned data sets will be the next logical step to eventually be able to automatically register multiple data sets in space and time.

\end{document}